\documentclass[usenatbib,fleqn]{mn2e} 
\usepackage{epsfig}
\usepackage{times}
\usepackage{fixltx2e} 
\usepackage{amsmath,amssymb}

\newcommand{\Msol}{{\,\rm M}_\odot} 
\newcommand{\kpc} {{\,\rm kpc}} 
\newcommand{\pc} {{\,\rm pc}}

\def\Myr{\,{\rm Myr}}

\newcommand{\K} {{\,\rm K}} 
\newcommand{\cc}{{\,\rm {cm^{-3}}}}
 
\newcommand{\kmsec}{{\,\rm {km\,s^{-1}} }}

\newcommand{\vel}{\mbox{$v$}}

\title[Gravitational stability of turbulent galactic discs] {Characterizing gravitational instability in turbulent multi-component galactic discs}

\author[Oscar Agertz, Alessandro B. Romeo and Kearn Grisdale] 
{Oscar Agertz$^{1}$\thanks{\tt o.agertz@surrey.ac.uk}, 
Alessandro B. Romeo$^{2}$ 
and Kearn Grisdale$^{1}$\\
$^1$Department of Physics, 
University of Surrey, 
Guildford, 
GU2 7XH, 
United Kingdom\\  
$^{2}$Department of Earth and Space Sciences, Chalmers University of Technology, SE-41296 Gothenburg, Sweden}
              
\date{\today}
\begin{document}
\maketitle

\begin{abstract}
Gravitational instabilities play an important role in galaxy evolution and in shaping the interstellar medium (ISM). The ISM is observed to be highly turbulent, meaning that observables like the gas surface density and velocity dispersion depend on the size of the region over which they are measured. In this work we investigate, using simulations of Milky Way-like disc galaxies with a resolution of $\sim 9 \pc$, the nature of turbulence in the ISM and how this affects the gravitational stability of galaxies. By accounting for the measured average turbulent scalings of the density and velocity fields in the stability analysis, we can more robustly characterize the average level of stability of the galaxies as a function of scale, and in a straightforward manner identify scales prone to fragmentation. Furthermore, we find that the stability of a disc with feedback-driven turbulence can be well described by a ``Toomre-like'' $Q$ stability criterion on all scales, whereas the classical $Q$ can formally lose its meaning on small scales if violent disc instabilities occur in models lacking pressure support from stellar feedback. 
\end{abstract}


\section{Introduction}
\label{sect:intro}
Today, over three decades after the pioneering work by \cite{larson81}, observations and
simulations of the interstellar medium (ISM) are revealing its turbulent
nature with higher and higher fidelity \citep[see review by e.g.][]{maclow:review04,elmegreen04}. Turbulence is not only thought to play an important role in controlling star formation in molecular clouds \citep{mckeeostriker07,PadoanNordlund2011,Federrath2012}, but also on galactic scales \citep{Renaud2013} and in shaping the interstellar medium (ISM) \citep{Stanimirovic1999,ElmegreenLMC2001,Bournaud2010}. The importance of galactic scale turbulence has in the past decade also been revealed in the early Universe; in gas rich high redshift galaxies, the observed levels of gas turbulence are much higher than in the local Universe \citep{Shapiro08,Forster2009,Swinbank2011}, explaining the observed ubiquity of super massive star forming clumps \citep{Elmegreen09,Agertz09b,Bournaud09,DSC09,Genzel2011}.

One of many fundamental aspects of ISM turbulence is the existence of scaling relations between observables, such as the column density ($\Sigma$), the 1D velocity dispersion ($\sigma$), and the size of
the region  ($\ell$) over which such quantities are measured:
\begin{equation}
\Sigma\propto\ell^{a},\;\;\;
\sigma\propto\ell^{b}.
\end{equation}
The values of exponents $a$ and $b$ depend on which ISM component and the range of scales that are considered. In this work we focus mainly on the cold neutral gas; neutral hydrogen (HI), and molecular gas, dominated by molecular hydrogen (H$_2$, observed via the tracer molecule CO), which is known to be supersonically turbulent and plays an important role in the gravitational instability
of galactic discs \citep[e.g.][]{LinShu1966,JogSolomon1984,bertinromeo88}.
 
In molecular gas, the scaling exponents are $a\approx0$ and $b\approx\frac{1}{2}$, up to scales of several 100 pc\footnote{Note that $a=0$ is expected for isolated clouds in gravitational equilibrium, as the cloud mass $M\propto \ell\sigma^2$ together with $\sigma\propto \ell^{0.5}$ gives $\Sigma\sim M/\ell^2= constant$.}. This pair of exponents are often referred to as ``Larson's scaling laws" after the discovery by \cite{larson81} \citep[see also][]{Solomon1987}. In fact, both Galactic and extragalactic Giant Molecular Clouds (GMCs) are
fairly well described by Larson's scaling laws, although with large uncertainties \citep[e.g.][]{bolatto_etal08,Heyer2009,Hughes2010,Kauffmann2010,Lombardi2010,Sanchez2010,Roman-Duval2010, Ballesteros-Paredes2011,Beaumont2012}, as well as in the dense star-forming clumps in high redshift galaxies \citep{Swinbank2011}.

Simulations of GMCs forming in galactic discs have recently started producing scaling relations compatible with Larson's relations \citep[e.g.][]{Hopkins2012structure,Dobbs2014}, but see \cite{Fujimoto2014} for models predicting steeper relations ($\Sigma(\ell),\sigma(\ell)\propto \ell$). State-of-the-art numerical simulations of supersonic turbulence suggest $a\sim\frac{1}{2}$ and $b\sim\frac{1}{2}$ \citep[][]{Fleck1996,Kowal2007,kritsuk07,Schmidt2008,PriceFederrath2010,Kritsuk2013}. Other recent numerical work suggest that the scaling exponent $a$ may be significantly affected by the nature of the turbulence forcing \citep{Federrath2010}, magnetic fields, and self-gravity \citep{Collins2012}.

In H\,\textsc{i}, the scaling exponents are $a\sim\frac{1}{3}$ and $b\sim\frac{1}{3}$ up to several kpc \citep{Roy2008}, although, as for molecular gas, with large uncertainties \citep{Kim2007}. Observed power spectra of H\,\textsc{i} intensity fluctuations in nearby galaxies are compatible with a Kolmogorov scaling for both $\sigma$ and $\Sigma$ \citep[e.g.][]{Stanimirovic1999, LazarianPogosyan2000,ElmegreenLMC2001,Begum2006,Bournaud2010,Zhang2012, Dutta2013}, with similar results for other ISM components on large scales \citep[e.g. dust and CO in M33, ][]{Combes2012}. Furthermore, H\,\textsc{i} power spectra are often found to be shallower on large scales, with a break around  $\ell\sim$ the disc scale height, possibly indicating a transition from 3D turbulence on small scales to 2D turbulence on large \citep[][]{Dutta2008,Dutta2009,Block2010,Bournaud2010}.

While the turbulent nature of the ISM is well established, it is rarely accounted for in theoretical works when evaluating the gravitational stability of galactic discs. Instead, $\sigma$ and $\Sigma$ are associated with smoothed quantities on galactic scales ($\sim$ kilo-parsecs) \citep[but see][]{Elmegreen1996,Begelman2009,HopkinsGMC2012}. A few analytical studies investigating the effect of ISM turbulence on gravitational stability have been carried out. Romeo et al.\ (2010) explored a range of values for $a$ and $b$ and showed that turbulence has an important effect on the gravitational instability of the disc; it excites a rich variety of stability regimes, several of which have no classical counterpart. Followup studies by \cite{HoffmannRomeo2012} and \cite{RomeoFalstad2013} \citep[see also][]{Shadmehri2012} extended this framework to turbulent multi-component (gas+stars) discs and applied it to the THINGS galaxy sample \citep{Walter2008}. Their analysis showed that H$_2$ plays a significant role in disc (in)stability by dynamically decoupling and dominating the onset for gravitational instability even at distances as large as half the optical radius.

The goal of this work is to in greater details explore the interplay between gas turbulence and disc stability by extending previous work in a number of ways:
\begin{enumerate}
\item We perform numerical simulations of a Milky Way-like galaxy with a resolution of $\sim 9\pc$, where we model the same galaxy in two markedly different ways: (1) without any stellar feedback, leading to rapid gas fragmentation into a population of star forming giant molecular clouds (GMCs), and (2) with efficient stellar feedback which acts to disperse the GMCs, drives interstellar turbulence, and regulates the rate of star formation. These two models show two extremes of galaxy evolution, and are useful platforms on which to understand, and characterize, the role of turbulence in realistic disc galaxies.
\item We apply a classical stability analysis on kpc-scales to both models and demonstrate how it fails to qualitatively separate the two systems, despite the dramatically different ISM morphology. We show that the this is in agreement with observations of nearby spiral galaxies.
\item By accounting for the scale-dependent nature of $\Sigma$ and $\sigma$ in a multi-component stability analysis, we demonstrate how we can more robustly characterize the disc stability and dynamical state of the galaxies, 
\end{enumerate}

The rest of the paper is organized as follows.  The classical framework for understanding the stability of
two-component thick discs is outlined in Section \,\ref{sect:gravinst} and \ref{sect:turb}. In Section \ref{sect:numtechnique} we describe the numerical hydro+$N$-body method used for the galaxy simulations. In Section \ref{sect:classical} we perform a classical analysis followed by an accounting of turbulent scaling relations in Section \ref{sect:turbanalysis}. In Section \ref{sect:mappingout} we discuss our results in a more general context of observed, and theoretically predicted, ISM scaling relations and what this means for understanding disc galaxies. Finally, we discuss and conclude our results in Section \ref{sect:discussion}.

\section{Method}
\label{sect:method}
\subsection{Stability diagnostics}
\label{sect:gravinst}
Consider a gas disc of scale height $h$, and perturb it with axisymmetric waves of frequency $\omega$ and wavenumber $k$.  The response of the disc is
described by the dispersion relation
\begin{equation}
\label{eq:disprel1}
\omega^{2}=\kappa^{2}-\frac{2\pi G\Sigma\,k}{1+kh}+\sigma^{2}\,k^{2}\,,
\end{equation}
where $\kappa$ is the epicyclic frequency, $\Sigma$ is the surface density at
equilibrium, and $\sigma$ is the sound speed \citep[][]{romeo92,romeo94} \citep[see also][]{Vandervoort1970}. The three terms on the right side of Eq.\, \ref{eq:disprel1} represent the contributions of rotation, self-gravity and pressure.  For $kh\ll1$, Eq.\ (1) reduces to the usual dispersion relation for an
infinitesimally thin gas disc \citep[for a derivation, see e.g.][]{BinneyTremaine2008}. 
Such a disc is unstable if and only if $\omega^2<0$, which is equivalent to $Q<1$, where $Q$ is defined by
\begin{equation}
\label{eq:toomreQ}
Q=\frac{\kappa\sigma}{\pi G \Sigma}.
\end{equation}
This quantity was first derived by \cite{toomre64} for a thin disc of stars (where $\sigma\rightarrow\sigma_R$, i.e. the stellar radial velocity dispersion, and the denominator has $\pi$ replaced by 3.36), and we hence refer to it as Toomre's $Q$. From now on we denote Eq.\,\ref{eq:toomreQ} for stars and gas as $Q_\star$ and $Q_{\mathrm{g}}$ respectively. For $kh\gg1$, one recovers the case of Jeans instability with rotation, since
$\Sigma/h\approx2\rho$.  In other words, scales comparable to $h$ mark the
transition from 2D to 3D stability.  

Much work has gone into characterizing gravitational instabilities in multi-component systems \citep[][]{bertinromeo88,romeo92,romeo94,Elmegreen1995,Jog1996,rafikov01}. Such systems are always more unstable than each component considered separately, and the interplay between the different components depends on the ratio between their velocity dispersions and their surface densities. \cite{RomeoWiegert2011} introduced a simple and accurate approximation for the two-component $Q$ parameter, which takes into account the stabilizing effect of disc thickness and predicts whether the local stability level is dominated by stars or gas. \cite{RomeoFalstad2013} generalized this approximation to discs made of several stellar and/or gaseous components, and to the whole range of velocity dispersion anisotropies observed in galactic discs.  In the two-component case, the $Q$ stability parameter is given by

\begin{eqnarray}
\frac{1}{Q_{\mathrm{thick}}}&=&
\left\{\begin{array}{ll}
       {\displaystyle\frac{W}{T_{\star}Q_{\star}}+
                     \frac{1}{T_{\mathrm{g}}Q_{\mathrm{g}}}}
                       & \mbox{if\ \ }T_{\star}Q_{\star}\geq
                                      T_{\mathrm{g}}Q_{\mathrm{g}}\,, \\
                       &                                              \\
       {\displaystyle\frac{1}{T_{\star}Q_{\star}}+
                     \frac{W}{T_{\mathrm{g}}Q_{\mathrm{g}}}}
                       & \mbox{if\ \ }T_{\mathrm{g}}Q_{\mathrm{g}}\geq
                                      T_{\star}Q_{\star}\,;
       \end{array}
\right.
\\
W&=&
\frac{2\sigma_{\star}\sigma_{\mathrm{g}}}
     {\sigma_{\star}^{2}+\sigma_{\mathrm{g}}^{2}}\,.
\\
T&\approx&
\left\{\begin{array}{ll}
       {\displaystyle1+0.6\left(\frac{\sigma_{z}}{\sigma_{R}}\right)^{2}}
                       & \mbox{for\ }0\la\sigma_{z}/\sigma_{R}\la0.5\,, \\
                       &                                                \\
       {\displaystyle0.8+0.7\left(\frac{\sigma_{z}}{\sigma_{R}}\right)}
                       & \mbox{for\ }0.5\la\sigma_{z}/\sigma_{R}\la1\,.
       \end{array}
\right.
\end{eqnarray}

The thin-disc limit, $Q_{\mathrm{thin}}$, can be recovered by setting $\sigma_{z}/\sigma_{R}=0$ and hence $T=1$. As for the single component case, the criterion for instability is  $Q_{\mathrm{thick}}<1$ and $Q_{\mathrm{thin}}<1$.

\subsection{Accounting for turbulent scalings}
\label{sect:turb}
As discussed in Sect. 1, observations and theory have revealed that the interstellar medium (ISM) is highly turbulent, and many properties of the ISM depend on the physical scale on which they are measured. As argued by \cite{Romeo2010}, turbulence can be accounted for in the stability diagnostics by generalizing the dispersion relation in Eq.\,\ref{eq:disprel1} to account for scale-dependencies of the velocity dispersion $\sigma$ and surface density $\Sigma$, i.e.
\begin{equation}
\label{eq:disprel2}
\omega^{2}=\kappa^{2}-2\pi G\Sigma(k)\,k+\sigma^{2}(k)\,k^{2},
\end{equation}
where $\Sigma(k)$ and $\sigma(k)$ are the mass column density and the 1D velocity dispersion (accounting both for the thermal and turbulent component) measured over a region of size $\ell=2\pi/k$. Observations and theoretical studies \citep[see e.g.][]{larson81,elmegreen04,mckeeostriker07,kritsuk07,Romeo2010} indicate a power-law behaviour in these quantities, motivating the following parametrization:
\begin{equation}
\label{eq:powerlaws}
\Sigma(k)=\Sigma_{0}\left(\frac{k}{k_{0}}\right)^{-a},\;\;\;
\sigma(k)=\sigma_{0}\left(\frac{k}{k_{0}}\right)^{-b}.
\end{equation}
If the disc has volume density $\rho$ and scale height $h$, then
$\Sigma\approx2\rho\ell$ for $\ell\la h$ and $\Sigma\approx2\rho h$ for
$\ell\ga h$.  The range $\ell\la h$ corresponds to the case of 3D turbulence
(relevant for GMCs and H\,\textsc{i} on small scales), whereas the range $\ell\ga h$ corresponds to the case of 2D turbulence (relevant for H\,\textsc{i} on large scales). The quantity $\ell_{0}=2\pi/k_{0}$ is the typical scale
at which $\Sigma$ and $\sigma$ are observed and quantities like the Toomre parameter $Q$ is computed, so that $Q_{0}=\kappa\sigma_{0}/\pi G\Sigma_{0}$. For example, at an angular resolution of $6^{\prime\prime}$ achieved for the "The HI Nearby Galaxy Survey" (THINGS) galaxy sample \citep{Walter2008}, analysis is carried out on the spatial smoothing scale $\ell_{0}\sim 0.5-1\kpc$ \citep{Leroy08}.

As we are interested in understanding the influence of turbulence on a two component disc (gas and stars), we need to analyze the joint dispersion relation which can then be expressed in the following form \citep[e.g.][]{JogSolomon1984,HoffmannRomeo2012}:
\begin{equation}
\label{eq:multi1}
\left(\omega^{2}-\mathcal{M}_{1}^{2}\right)
\left(\omega^{2}-\mathcal{M}_{2}^{2}\right)=
\left(\mathcal{P}_{1}^{2}-\mathcal{M}_{1}^{2}\right)
\left(\mathcal{P}_{2}^{2}-\mathcal{M}_{2}^{2}\right),
\end{equation}
where the (turbulent) dispersion relation for each component $i$ has the usual form 
\begin{equation}
\label{eq:multi2}
\mathcal{M}_{i}^{2}\equiv
\kappa^{2}-2\pi G\Sigma_{i}(k)\,k+\sigma_{i}^{2}(k)\,k^{2},
\end{equation}
and
\begin{equation}
\label{eq:multi3}
\mathcal{P}_{i}^{2}\equiv
\kappa^{2}+\sigma_{i}^{2}(k)\,k^{2},
\end{equation}
with $i=$gas, star\footnote{The dispersion relation of an $N$-component turbulent disc is  $\sum_{i=1}^{N} (\mathcal{M}_{i}^{2}-\mathcal{P}_{i}^{2}) /
  (\omega^{2}-\mathcal{P}_{i}^{2}) = 1$, as can be inferred from
  equaiton 22 of Rafikov (2001).}.
Note that $\mathcal{M}_{i}^{2}(k)$ is the one-component dispersion
relation for potential-density waves, while
$\mathcal{P}_{i}^{2}(k)$ describes sound waves modified by rotation and
turbulence. Furthermore, the right-hand side of Eq.\,\ref{eq:multi1} measures the strength of the gravitational coupling between the two components, as $\mathcal{M}_{i}^{2}(k)-\mathcal{P}_{i}^{2}(k)$ is the self-gravity term of component $i$. Eq.\,\ref{eq:multi1} is quadratic in $\omega^2$, with two real roots:
\begin{equation}
\label{eq:coupleddisp}
\omega_{\pm}^{2}=
\frac{1}{2}\left[\mathcal{M}_{1}^{2}+\mathcal{M}_{2}^{2}
\pm\sqrt{\Delta}\,\right],
\end{equation}
\begin{equation}
\Delta=
\left(\mathcal{M}_{1}^{2}-\mathcal{M}_{2}^{2}\right)^{2}+4
\left(\mathcal{P}_{1}^{2}-\mathcal{M}_{1}^{2}\right)
\left(\mathcal{P}_{2}^{2}-\mathcal{M}_{2}^{2}\right).
\end{equation}
This means that the dispersion relation has two branches that do not cross,
$\omega_{+}^{2}(k)\neq\omega_{-}^{2}(k)$, except possibly as $k\rightarrow0$
or $k\rightarrow\infty$. We focus only on $\omega_{-}^{2}(k)$ in this work, as 
this branch is the only one which can represent gravitational instability 
($\omega_{+}^{2}(k)$ is manifestly $\ge 0$). Note that $\omega_{-}^{2}(k)$ is always \emph{smaller} than each component individually.

In Section \ref{sect:turbanalysis} and \ref{sect:mappingout} we use the above framework, together with numerical simulations, to demonstrate how turbulence affects the shape of the dispersion relation, and hence the condition for local gravitational instability ($\omega^{2}<0$). 

\subsection{Numerical technique}
\label{sect:numtechnique}
In order to carry out hydro+$N$-body simulations of galactic discs, we use the Adaptive Mesh Refinement (AMR) code {\tt RAMSES} \citep{teyssier02}. The fluid dynamics of the baryons is calculated using a second-order unsplit Godunov method, while the collisionless dynamics of stellar and dark matter particles is evolved using the particle-mesh technique \citep{Hockney1981}, with gravitational accelerations computed from the gravitational potential on the mesh. The gravitational potential is calculated by solving the Poisson equation using the multi-grid method \citep{GuilletTeyssier2011} for all refinement levels. The equation of state of the fluid is that of an ideal mono-atomic gas with an adiabatic index $\gamma=5/3$. 

The code achieves high resolution in high density regions using adaptive mesh refinement, where the refinement strategy is based on a quasi-Lagrangian approach in which the number of collisionless particles per cell is kept approximately constant. This allows the local force softening to closely match the local mean interparticle separation, which suppress discreteness effects \citep[e.g.,][]{Romeo08}. An analogous refinement criterion is also used for the gas.

The star formation, cooling physics and stellar feedback model adopted in our simulations is described in detail in \cite{Agertz2013} (identical to the "{\tt All}" model) and \cite{AgertzKravtsov2014}, and we refer the reader to those papers for details. Briefly, several processes contribute to the stellar feedback budget, as stars inject energy, momentum, mass and heavy elements over time via SNII and SNIa explosions, stellar winds and radiation pressure into the surrounding gas. 
Metals injected by supernovae and stellar winds are advected as a passive scalar and are incorporated self-consistently in the cooling and heating routine. 
 
One aspect differs from previous work; at the current numerical resolution ($\Delta x\sim 9\pc$), we are not guaranteed to always resolve the cooling radius\footnote{the cooling radius scales as $r_{\rm cool}\approx 30 n_0^{-0.43}(Z/Z_\odot+0.01)^{-0.18}$ pc for a supernova explosion with energy $E_{\rm SN}=10^{51}$ erg \citep[e.g.][]{Cioffi1988,Thornton1998,KimOstriker2014}} $r_{\rm cool}$, i.e. the SN bubble radius for which radiative losses are expected to be important for each discrete SN event, leading to an underestimation of the impact of SNe feedback. Instead of remedying this issue by solving two separate energy equations, one for the thermal energy and one for the feedback energy as in \cite{AgertzKravtsov2014}, we adopt the model recently suggested by \cite{KimOstriker2014} \citep[see also][]{Martizzi2014,Gatto2014,Simpson2014}. Here a SN explosion resolved by at least three grid cells ($r_{\rm cool}\geq3\Delta x$) is initialized in the energy conserving phase by injecting the relevant energy ($10^{51}\,$erg per SN) into the nearest grid cell. If this criterion is not fulfilled, the SN is initialized in its momentum conserving phase, i.e. the momentum generated during the energy conserving Sedov-Taylor phase is injected into to the 26 cells surrounding a star particle. 

\begin{figure*}
\begin{center}
\includegraphics[width=0.75\textwidth]{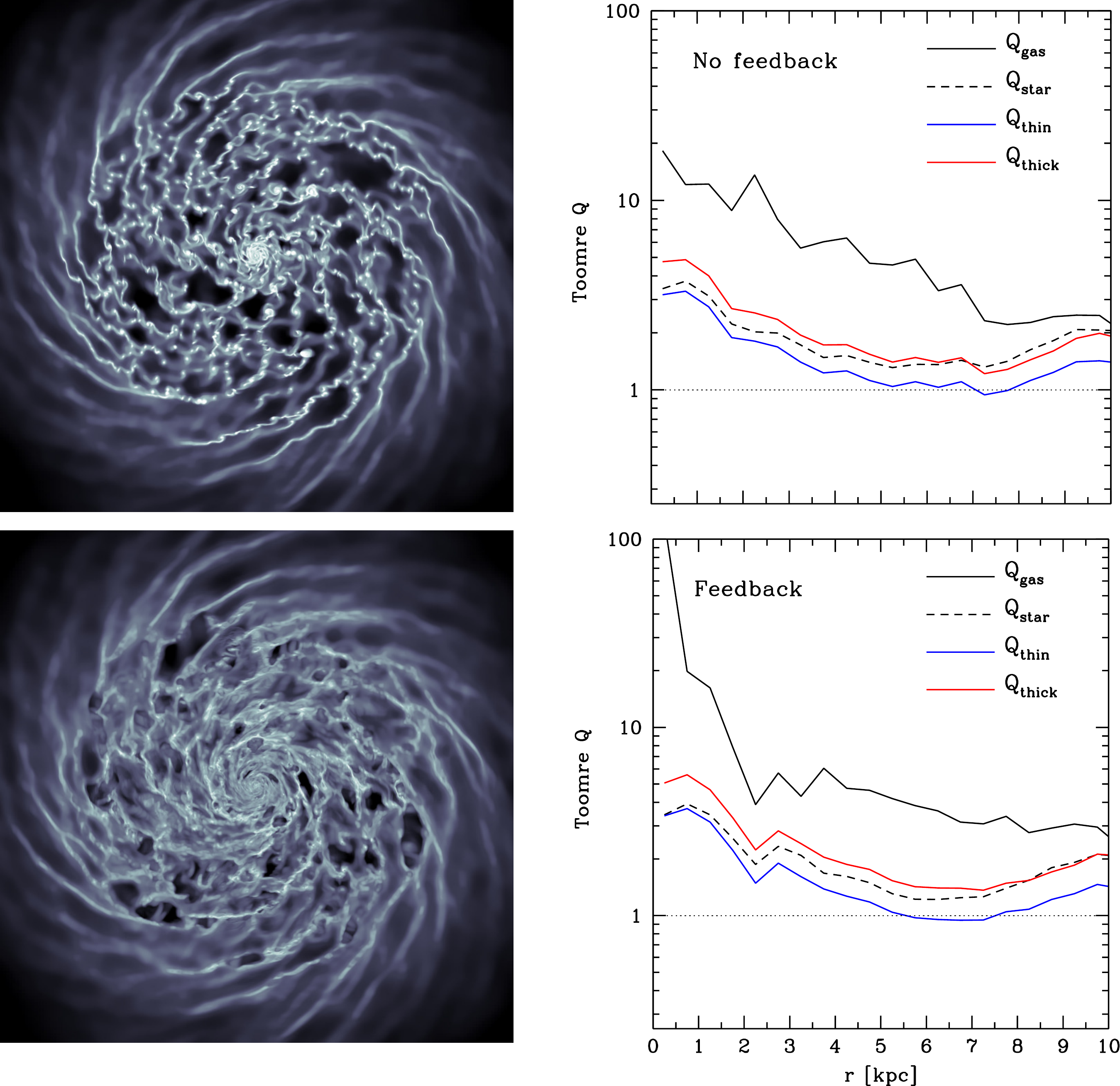}
\caption{(Left) Projected gas density maps covering a region $36\times36\kpc$ in size at $t=140$ Myr of the simulation without (top) and with (bottom) stellar feedback. ({\it Right}) Mass weighted average profiles, computed in radial bins of size $\Delta r=0.5\kpc$ of the classical Toomre $Q$ ($=\sigma\kappa/\pi G\Sigma$) for gas and stars, the joint stability parameter $Q_{\rm thin}$ and the joint parameter accounting for disc thickness $Q_{\rm thick}$ (Romeo \& Wiegert 2011; Romeo \& Falstad 2013). 
}
\label{fig:Q}
\end{center}
\end{figure*}

\cite{Blondin1998} calculated the transition time from the energy conserving phase to the phase of shell formation, at which the cooling time equals the age of the remnant ($t_{\rm cool}=t_{\rm SN}$), to be $\approx 2.9\times 10^4\,E_{51}^{4/17}n_0^{-9/17}\,{\rm yrs}$, where $n_0$ is the ambient density and $E_{51}$ the thermal energy in units of $10^{51}~{\rm ergs}$. At this time, the momentum of the expanding shell is approximately
\begin{equation}
\label{eq:ST}
p_{\rm ST}\approx 2.6\times 10^5\,E_{51}^{16/17}n_0^{-2/17} \Msol\kmsec.
\end{equation}
For reasonable values of ambient densities, this is $\sim 10$ times greater than the initial ejecta momentum. \cite{KimOstriker2014} and \cite{Martizzi2014} have shown, using detailed simulations of SNe explosions, that Eq.\,\ref{eq:ST} holds even for more realistic, clumpy, environments. We hence use this relation for the injected momentum per individual SN explosion when the cooling radius is not resolved.

\subsubsection{Simulations}
\label{sect:simulations}
We model the non-linear evolution of an entire Milky Way-like galactic disc. This setup is now the standardized test for the {\tt AGORA} code comparison project \citep{agora}, and is a higher resolution version of the galaxy analyzed in \cite{Agertz2013}. Briefly, following \cite{Hernquist1993} and \cite{Springel2000} \citep[see also][]{SpringelMatteoHernquist2005} we create a particle distribution representing a late type, star forming spiral galaxy embedded in an NFW dark matter halo \citep{nfw1996}. The dark matter halo has a concentration parameter $c=10$ and virial circular velocity, measured at the overdensity $200\rho_{\rm crit}$, $\vel_{\rm 200}=150\kmsec$, which translates to a halo virial mass $M_{\rm 200}=1.1\times 10^{12}\Msol$. The total baryonic disc mass is $M_{\rm disc}=4.5\times10^{10}\Msol$ with $20\%$ in gas. The bulge-to-disc mass ratio is $B/D=0.1$. We assume exponential profiles for the stellar and gaseous components and adopt a disc scale length $r_{\rm d}=3.4\kpc$ and scale height $h=0.1r_{\rm d}$ for both. The bulge mass profile is that of \cite{Hernquist1990} with scale-length $a=0.1r_{\rm d}$. The halo and stellar disk are represented by $10^6$ particles each, and the bulge consists of $10^5$ particles.
 
We initialize the gaseous disc analytically on the AMR grid assuming an exponential profile. The galaxy is embedded in a hot ($T=10^6\K$), tenuous ($n=10^{-5}\cc$) gas halo enriched to $Z=10^{-2}Z_\odot$, while the disc has solar abundance. The minimum AMR cell size reached in the simulations is $\Delta x=9\pc$. 

We carry out two simulations, one without any stellar feedback, adopting a local star formation efficiency per free-fall time \citep[see e.g.][]{Agertz2013} $\epsilon_{\rm ff}=1\%$, and one with stellar feedback but with $\epsilon_{\rm ff}=10\%$. In the former case, the effect of feedback is implicit in the choice of $\epsilon_{\rm ff}$, and in the latter case this is achieved by efficient stellar feedback. Both simulations therefor have similar star formation histories and normalizations of the $\Sigma_{\rm SFR}-\Sigma_{\rm gas}$ (Kennicutt-Schmidt) relation \citep[see discussion in][]{Agertz2013,AgertzKravtsov2014}, but reach this state in different ways. 

\begin{figure*}
\begin{center}
\includegraphics[width=0.72\textwidth]{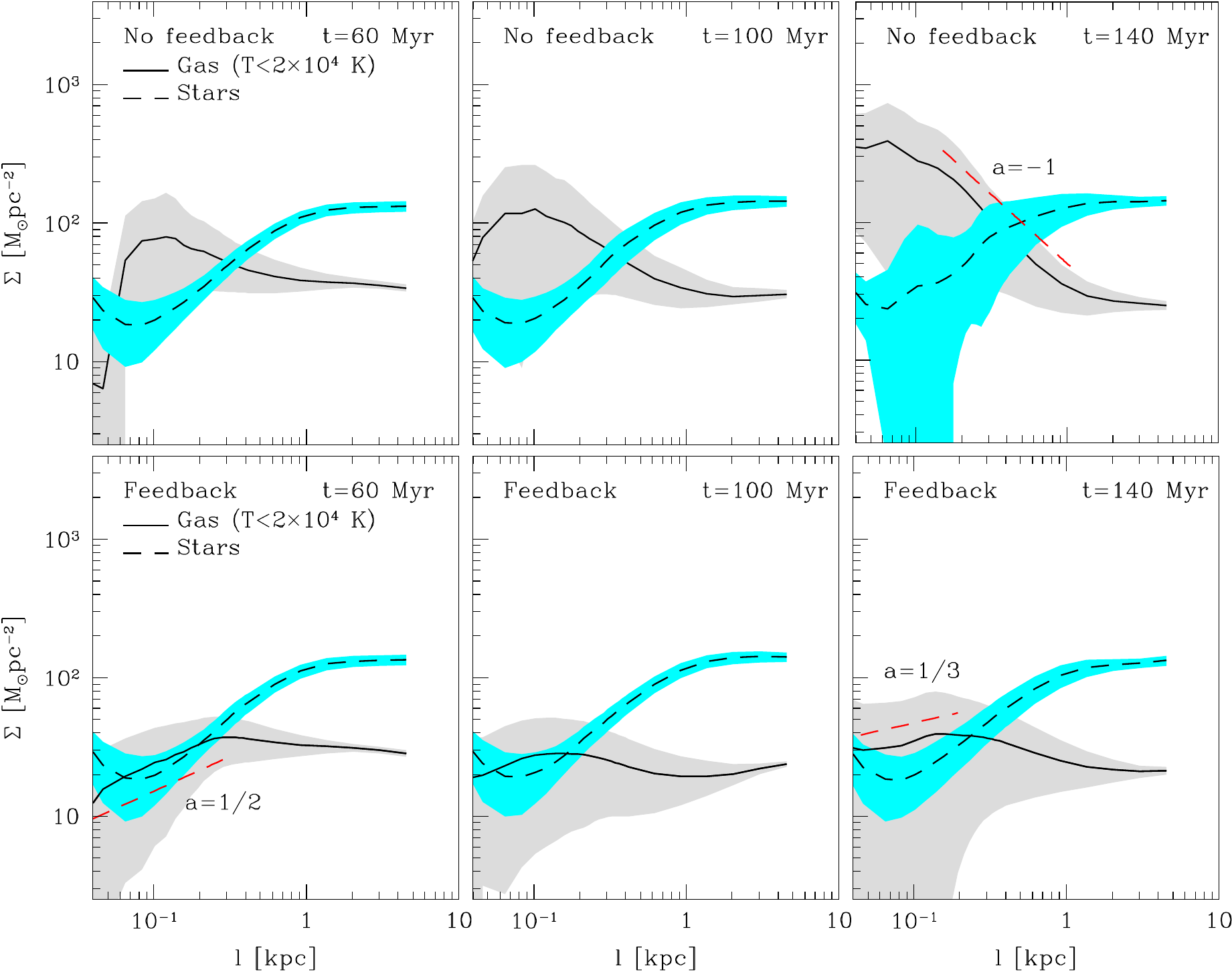}
\caption{The scale dependence of the gas and stellar surface density $\Sigma$ in the models adopting no feedback (top) and feedback (bottom) for, from left to right, $t=60,100$ and 140 Myr.}
\label{fig:scaling}
\end{center}
\end{figure*}
\begin{figure*}
\begin{center}
\includegraphics[width=0.72\textwidth]{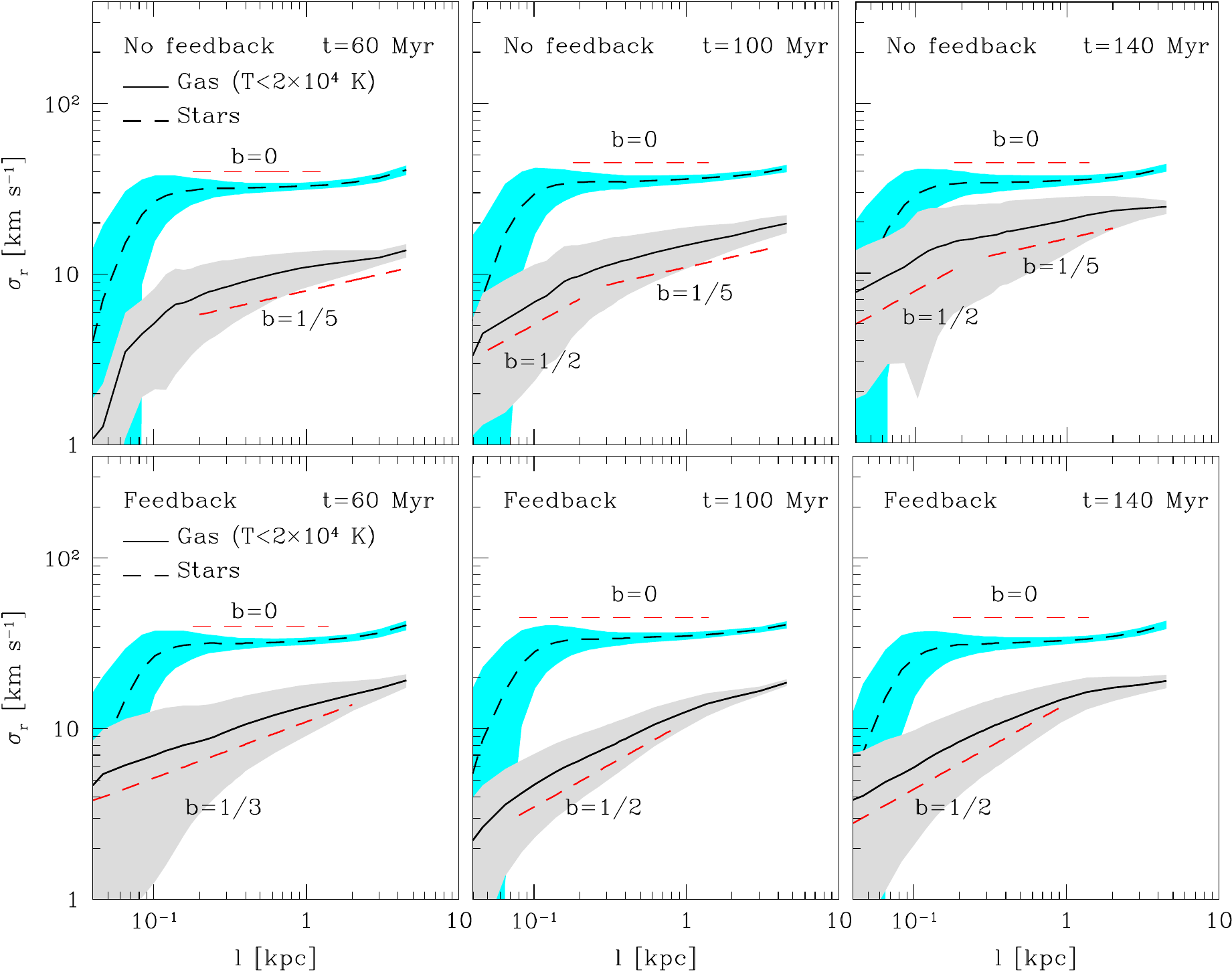}
\caption{The scale dependence of the gas and stellar radial velocity dispersion $\sigma_R$ in the models adopting no feedback (top) and feedback (bottom) for, from left to right, $t=60,100$ and 140 Myr.
}
\label{fig:scaling2}
\end{center}
\end{figure*}

\section{Results}
\subsection{Classical stability analysis}
\label{sect:classical}
The left hand side of Fig.\,\ref{fig:Q} shows projected gas surface density maps of the two  simulated galactic discs at $t=140~\Myr$. Without stellar feedback, the disc violently fragments into a long-lived population of massive star forming GMCs, whereas the clouds are rapidly dispersed, and reformed, in the feedback regulated case. In the right hand panel we show radial profiles of the classical Toomre $Q$ for gas and stars, as well as the two-component thin and thick disc stability parameter, $Q_{\rm thin}$ and $Q_{\rm thick}$ respectively (see \S\,\ref{sect:gravinst}). 

We find all $Q$ radial profiles to be almost indistinguishable between the two models, and note that they are in agreement with derived values from well resolved spiral galaxies, see e.g. the detailed analysis by \cite{Leroy08} of the THINGS galaxies \citep[see also figure 5 in][]{RomeoWiegert2011,RomeoFalstad2013}; e.g. $Q_{\rm thick}\sim 1.5-3$ for $r\gtrsim 2\kpc$, i.e. outside the bulge, indicative of gravitational stability to axisymmetric waves. The similar values of $Q(r)$, all being significantly larger than unity, arise despite the morphological states of the discs being markedly different. 

\cite{Leroy08} concluded that the large observed values of $Q$, indicative of local stability\footnote{Note that this refers to stability against axisymmetric perturbations and that stability against \emph{non}-axisymmetric perturbations is not guaranteed \citep[e.g.][]{GrivGedalin2012}}, put doubts on the role of gravitational instabilities in \emph{alone} controlling the local star formation efficiency. However, as we have been alluding to in previous sections, our analysis is done on a fixed, and rather large, scale ($\ell_0 = 0.5 \kpc$) chosen to coincide with the spatial resolution of the THINGS survey. The fact that different indicators of stability fail to separate the simulated galaxies means that galactic dynamics, on the chosen smoothing scale, is the same. To probe differences further we need to adopt a scale dependent stability analysis.

\begin{figure*}
\begin{center}
\includegraphics[width=0.8\textwidth]{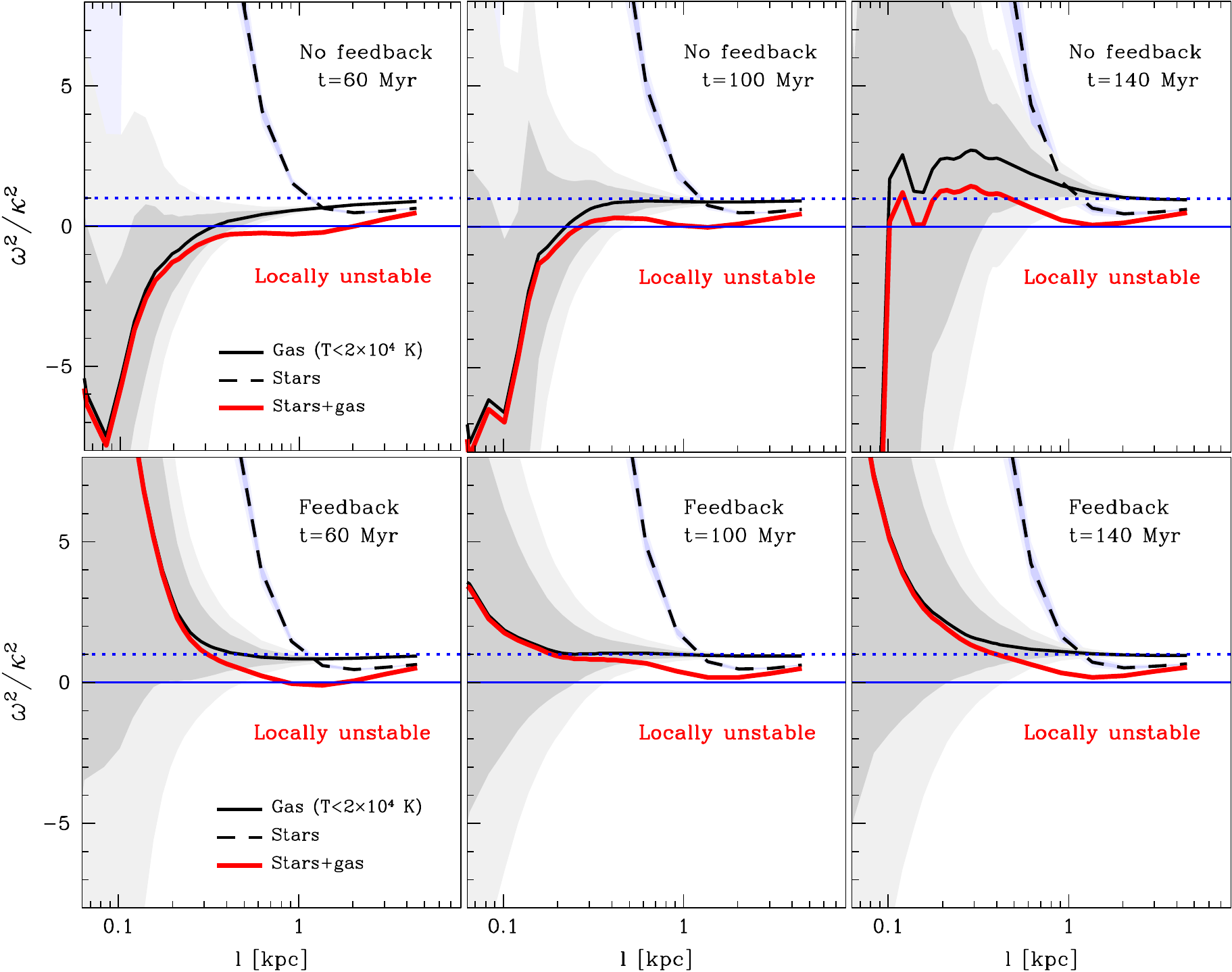}
\caption{The instability growth rate, accounting for scale dependent $\sigma(\ell)$ and $\Sigma(\ell)$, for gas (solid lines) and stars (dashed lines) for no feedback (top) and feedback (bottom). When turbulent scalings are accounted for in the analysis, the feedback simulation is found to be \emph{on average} stable on all scales, with a Toomre-like behaviour (see main text), whereas the model lacking feedback is characterized as unstable on small scales ($\ell\lesssim$ a few $100\,\pc$), in accordance with the fragmented gas morphology with star forming clouds of sizes $\sim 100$ pc.}
\label{fig:w2}
\end{center}
\end{figure*}

\subsection{Accounting for turbulent scalings}
\label{sect:turbanalysis}
In this section we characterize the effect of turbulence on the gravitational stability of the simulated discs directly via the dispersion relation for gas and stars (Eq.\,\ref{eq:disprel2}), as well as the relation for the coupled system ($\omega^2_{-}$ in Eq.\,\ref{eq:coupleddisp}).

We estimate $\Sigma(\ell)$ and $\sigma(\ell)$ for gas and stars from the simulations at different simulation times as follows; assume the disc plane coincides with the $xy$-plane, i.e. the axis of rotation is parallel with the $z$-axis. 
Centered in the mid-plane of the disc ($z=0$), we place cubes in a lattice configuration in the range $(x,y)\in\{-15,15\}\kpc$, with a lattice spacing of $\Delta l=100\pc$. Starting from a cube size of $\ell=36\pc$, we incrementally increase this value by $2\Delta x=36\pc$ up to $\ell=0.5\kpc$, after which we increase the cube sizes in steps of $50 \%$ of the current value up to $\ell\sim10\kpc$, simply to reduce the computational cost. On each scale we compute $\Sigma_i(\ell)$ and $\sigma_i(\ell)$ for each component (gas and stars) for every cube $i$. We then define $\Sigma(\ell)$ and $\sigma(\ell)$ to be the averages of all cubes at any given scale, and consider these quantities to be representative of the typical surface density and velocity dispersion. We have confirmed that sampling the discs in different way, e.g. randomly, has a negligible effect on the results presented below.

From now on we focus our analysis on the average scale dependent characteristics centered in an azimuth defined by $4\kpc<r<5\kpc$. Furthermore, as the gaseous velocity dispersion is found not be isotropic \citep[see also discussion in][]{Agertz09}, we will from here on consider the radial velocity dispersion $\sigma_R(\ell)$ for both gas and stars, as this is the relevant component entering the dispersion relation for non-isotropic velocity ellipsoids. Note also that the sound speed of the gas $c_s\ll\sigma_R$ in a mass weighted sense; the cold ISM is dominated by super-sonic motions, and we can in principle omit $c_s$ in the stability analysis.

\subsubsection{Density evolution and spatial scaling}
Fig.\,\ref{fig:scaling} shows the time evolution ($t=60,100$ and 140 Myr) of $\Sigma_{\rm gas}(\ell)$ and $\Sigma_{\rm star}(\ell)$ for the fragmenting (no feedback) and feedback regulated discs. Both simulations show ``saturated" quantities above $\ell \sim 1\kpc$, above which little variation with scale is found, confirming the strikingly similar $Q(r)$ values found in \S\,\ref{sect:classical}. On smaller scales, the two models evolve differently; without feedback at $t=140\Myr$ (and as well for earlier times, but on scales of a few 100 pc), the population of bound star forming clouds, arising from violent fragmentation, leads to a steeply decreasing $\Sigma_{\rm gas}(\ell)$ with increasing $\ell$, while the feedback regulated case shows a rather flat $\Sigma_{\rm gas}(\ell)$. Quantified using the power-law relation in Eq.\,\ref{eq:powerlaws} ($\Sigma\propto \ell^{a}$), the clumpy galaxy approaches $a\sim-1$ and the feedback regulated $a\sim 0$ for $\ell\lesssim 1\kpc$. Significant scatter exists in the data, owing to the structured nature of the ISM and the fact that we do not bias the analysis to any specific regions of the disc\footnote{For example, by not enforcing the analysis to be centered \emph{only} on the dense cloud population in the fragmented disc, we do not measure a monotonically increasing $\Sigma(\ell)$ as $\ell\rightarrow 0$.} or phases of the gas, other than the cold ISM.
 
$\Sigma_{\rm star}(\ell)$ is increasing at scales larger than $\sim 100\pc$. This is due to the analysis being done in mid-plane centered cubes, for which scales $\ell\sim 1\kpc$ do not encapsulate all of the (kinematically hot) stars present in the disc. A similar conclusion may be drawn for the cold gas in the feedback regulated simulation, where the small scale turbulence, in an average sense, allows for a positive value of $a$. Adopting a scale dependent analysis for both stars and gas hence automatically accounts for disc thickness, without the need to adopt an effective surface density $\Sigma_{\rm eff}=\Sigma/(1+kh)$, as is done in Eq.\,\ref{eq:disprel1}. 

\subsubsection{Velocity dispersion evolution and spatial scaling}
\label{sect:sigmascaling}
Fig.\,\ref{fig:scaling2} shows the time evolution, $t=60,100$ and 140 Myr, of $\sigma_{R, {\rm star}}(\ell)$ and $\sigma_{R, {\rm gas}}(\ell)$ for the fragmenting (no feedback) and feedback regulated discs. The stars show a roughly scale independent value of $\sigma_R\sim 30\kmsec$ in both disc models. For the cold gas we find that the velocity dispersions are, after an initial transient, quite similar both in magnitude and dependency on scale regardless of the presence of feedback or not. In both models, and for $t>60\Myr$, $\sigma_{R, {\rm gas}}(\ell)\propto\ell^{1/2}$ on scales $\ell\lesssim 300-400\pc$ in the feedback model, and $\ell\lesssim 100\pc$ in the model without, in excellent agreement with the observed Larson-like scaling relevant for GMCs (see \S\,\ref{sect:intro}).

Still at times $t>60\Myr$, and on large scales (a few $100 \pc\lesssim\ell\lesssim 4\kpc$), the slope of $\sigma_{R, {\rm gas}}(\ell)$ transitions from $b\sim1/2$, into a flatter profile, with $b\sim1/5$, in the case without feedback. In the feedback regulated galaxy we measure $b\sim1/2$ up to almost $\sim 1\kpc$, with a flattening on scales  $\gtrsim1\kpc$. The slopes, and the scales over which the relation is measured in the feedback model, are in excellent agreement with observations of the Galactic HI linewidth--scale relation. For example, \cite{Kim2007} studied this relation for individual HI clouds in the Large Magellanic Cloud (LMC) and found $b\sim 0.3-0.5$ on scales $\lesssim 0.5-1\kpc$ (down to parsec scales).

We emphasize that both models show values of $\sigma\sim10-20\kmsec$ on large scales, in agreement with both HI and CO observations in local spiral galaxies \citep[][]{Tamburro2009, CalduPrimo2013}, despite the different nature in ISM turbulence driving on small scales. In the case of feedback-driven turbulence, star forming clouds are rapidly destroyed by internal process and the gas is dispersed, whereas disordered gas motions are driven by gravitational instabilities and galactic shear, leading to clump-clump interactions \citep[e.g.][]{jogostriker88,gammie91,Agertz09, tasker09} in the case without feedback. 

\subsubsection{The dispersion relation}
\label{sect:thedisprel}
Fig.\,\ref{fig:w2} shows the resulting mass weighted average dispersion relations, and the associated scatter, for the two galactic models for the stars ($\omega^2_{\rm stars}$), cold gas ($\omega^2_{\rm gas}$) and the coupled system ($\omega^2_{-}$). We remind the reader that the average $\omega^2(\ell)$ relations now self-consistently account for the actual scale dependent values of $\sigma(\ell)$ and $\Sigma(\ell)$ present in each analyzed region of the galaxies.

While featuring an almost identical $Q(r)$ (for $t=140\Myr$, \S\ref{sect:classical}), when analyzed on $\sim 1 \kpc$ scales, we can here identify the level of stability on all scales. Both discs are indeed stable or marginally stable ($\omega^2\geq0$) on large scales, and have in fact almost identical $\omega^2(\ell)$ relations for $\ell\gtrsim1\kpc$. 

The fragmented disc without stellar feedback always shows $\omega^2_{\rm gas}<0$ and $\omega^2_{-}<0$ on small scales, as suspected simply via visual inspection of the top left panel of Fig.\,\ref{fig:Q}. At early times ($t=60\Myr$), when fragmentation has just occured, the scale of instability is formally as large as $\sim 2\kpc$, with scales below a few $\sim 100\pc$ being the most unstable. At subsequent times, $\omega^2_{-}<0$ on scales $\ell\lesssim 100-200 \pc$, roughly the maximum size of GMCs, or more correctly GMAs (Giant Molecular Associations) in this model. This is also the scale at which we measured a transition in the Larson-like scaling of the gas velocity dispersions, from $b\sim1/2$ into $b\sim 1/5$ (\S\,\ref{sect:sigmascaling}). 

The above conclusions are in stark contrast to the feedback regulated case which is, at least on average, stable or marginally stable on all scales. Note that individual patches of the disc on small scales can be unstable, leading to star formation, as is evident for $\omega^2_{\rm gas}(\ell\lesssim 100\pc)$ at a $\gtrsim 0.5\,\sigma$ level. 

This analysis underscores the necessity to extend a traditional stability analysis with scale dependent variables to account for the typical average velocity and density structures that exists in the ISM. Furthermore, many classical concepts, such as a well defined fastest growing mode for a Toomre unstable ($Q<1$) disc, may no longer exist for the emerging scalings of $\Sigma(\ell)$ and $\sigma(\ell)$ introduced by strong fragmentation, as pointed out by \cite{Romeo2010} and \cite{Romeo2014}.

\begin{figure}
\begin{center}
\includegraphics[width=0.49\textwidth]{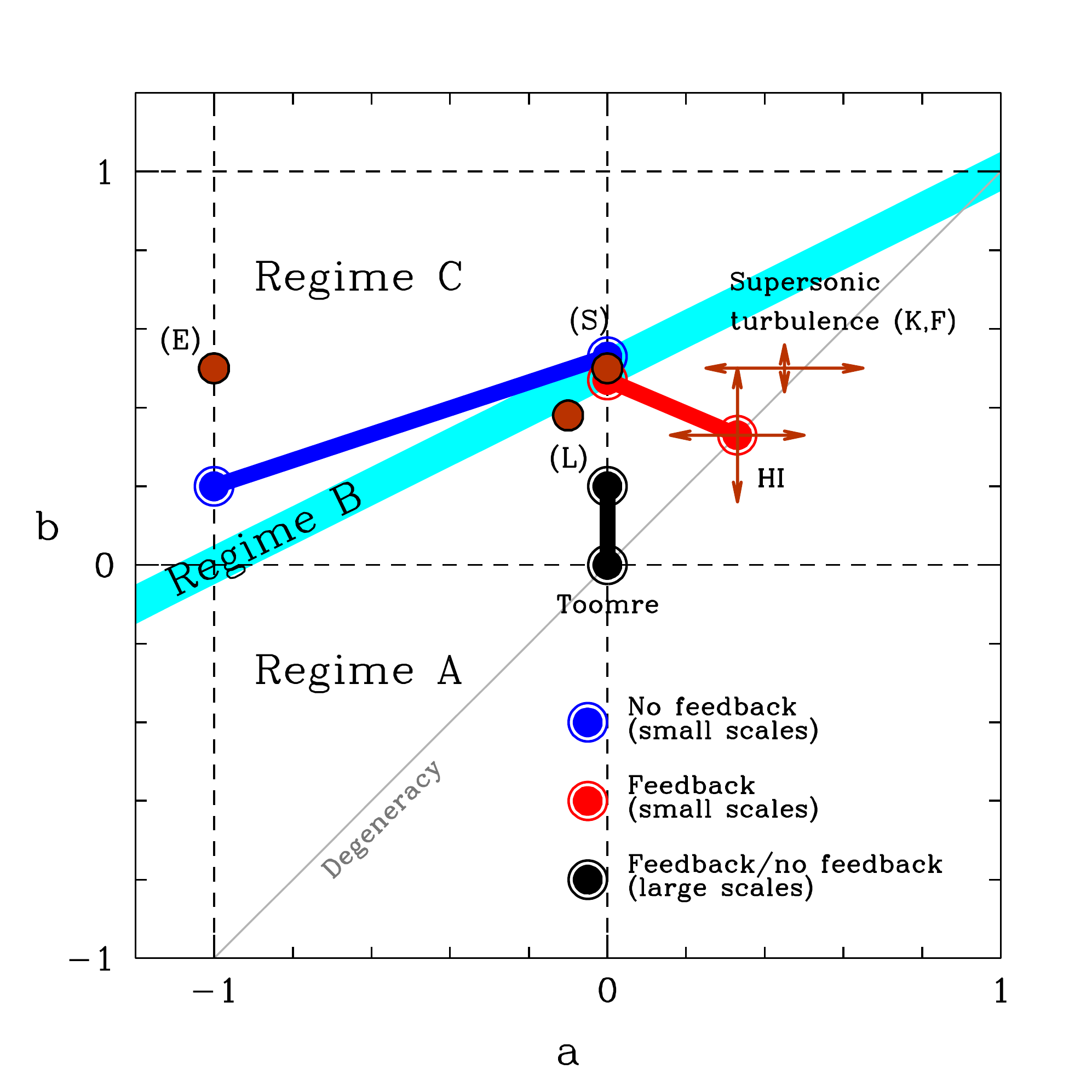}
\caption{The stability map of turbulence, with typical values derived from observations and numerical simulations, see the main text for a comprehensive discussion. The blue, red and black points indicate the regimes found in the simulated disc galaxies on both small and large scales. On large scales ($\ell\gtrsim 1\kpc$) the feedback regulated disc and the fragmented disc converge to roughly the same average $\Sigma$ and $\sigma$, and both well described by a Toomre-like stability criterion (Regime A, see text). Turbulent scaling on small scales pushes the fragmented disc into a regime typical observed for GMCs (Regime B), and a regime where the classical Toomre analysis is no longer valid (Regime C). The disc with feedback-driven turbulence is, in a statistical sense, stabilized on small scales, and shows scalings in broad agreement with HI and CO observations.
}
\label{fig:stabmap}
\end{center}
\end{figure}

\subsection{Mapping out the stability regimes}
\label{sect:mappingout}
How do the results in the previous sections connect to observed turbulent scaling in the ISM and results from high resolution numerical simulations? Fig.\,\ref{fig:stabmap} shows the \emph{stability map of turbulence} \citep{Romeo2010}, where the axes denote the $a$ and $b$ power-law exponents for the surface density and the velocity dispersion (Eq.\,\ref{eq:powerlaws}). 

Following the analysis in \cite{Romeo2010} \citep[see also][]{HoffmannRomeo2012}, we have identified, and indicated in the figure, three regimes of stability:

\begin{itemize}
\item{\it Regime A:} For $b<\frac{1}{2}\,(1+a)$ and $-2<a<1$, the stability of the disc is controlled by $Q_{0}=\kappa\sigma_{0}/\pi G\Sigma_{0}$ (i.e. the $Q$ measured on the fiducial smoothing scale $\ell_0$, see \S\,\ref{sect:turb}): the disc is stable at
  all scales if and only if $Q_{0}\geq\overline{Q}_{0}$, where
  $\overline{Q}_{0}$ depends on $a$, $b$ and $\ell_{0}$, and must be derived from Eq.\,\ref{eq:disprel2} for a single component system.  This is the domain
  of H\,\textsc{i} turbulence. Both H\,\textsc{i} observations and
  high-resolution simulations of supersonic turbulence are consistent with
  the scaling $a\sim b$.  In the case when $a=b$, the ratio $\sigma(\ell)/\Sigma(\ell)$ is a constant and the local stability criterion
  degenerates into the classical Toomre case, $Q_{0}\geq1$, as if the disc was non-turbulent and infinitesimally thin.
\item{\it Regime C:} For $b>\frac{1}{2}\,(1+a)$ and $-2<a<1$,
  the stability of the disc is no longer controlled by $Q_{0}$: the disc is
  always unstable on small scales (i.e.\ as $l\rightarrow 0$) and stable
  on large scales.
\item{\it Regime B:} For $b=\frac{1}{2}\,(1+a)$ and $-2<a<1$,
  the disc is in a phase of transition between Toomre-like stability (Regime A) and instability om small scales (Regime C).  This is the domain typically observed for molecular gas in GMCs. 
 \end{itemize}

A few well studied examples from theory/simulations and observations in the literature are shown in the figure:

\begin{itemize}
\item (L), $(a, b) = (-0.1, 0.38)$: The original scaling relations in GMCs found by \cite{larson81}.
\item (S), $(a, b) = (0, 0.50 \pm 0.05)$: The scaling relations in GMCs found by \cite{Solomon1987}. These values are what is usually referred to as ``Larson's scaling laws".
\item (K), $(a,b) = (\frac{1}{3}, \frac{1}{2})$: The result of high-resolution simulations of supersonic turbulence \citep[][Kritsuk, private communication]{Kritsuk2013}.
\item (F), $(a, b) = (0.44 \pm 0.14, 0.49 \pm 0.02)$: A prediction based on state-of-the-art simulations of super-sonic turbulence with compressive driving \citep[][Federrath, private communication]{Federrath2013}. For solenoidal driving, $(a, b) = (0.58 \pm 0.03, 0.48 \pm 0.02)$.
\item (E), $(a, b) = (-1, 0.5)$: Investigated by \cite{Elmegreen1996}
\item (HI): Typical range of values derived from observed HI intensity fluctuations in disc galaxies \citep[][]{LazarianPogosyan2000,Kim2007,Roy2008}.
\end{itemize} 

A number of typical $(a,b)$ pairs measured in the two simulations are indicated in Fig.\,\ref{fig:stabmap}. In the case of no feedback, leading to strong fragmentation, the disc features $a\sim -1$ to $0$, and $b\sim 1/5-1/2$ on small/intermediate scales. This puts the disc in regime B and C, with the latter meaning that the disc is always unstable on small scales. In fact, by studying the dispersion relations ($\omega^2_{\rm gas}(\ell)$ or $\omega^2_{-}(\ell)$) in Fig.\,\ref{fig:w2}, no well defined minimum exists, with the smallest numerically resolved scale, here on the order of a cell size $\sim \Delta x$, being the fastest growing mode. On large scales, we measure a transition into Regime A with $(a,b)\sim(0,0)$, i.e. the classical Toomre case with a well defined average surface density and velocity dispersion. 

The feedback driven simulation shows the same behaviour on large scales ($\ell\gtrsim1\kpc$) as the fragmented counterpart, coinciding with the classical Toomre case. On small scales ($\ell\lesssim1\kpc$), the simulations diverge, as discussed already in \S\,\ref{sect:thedisprel}. Here feedback creates an, on average, marginally stable ISM, still in regime A, with $a$ and $b$ values compatible with observations of the cold galactic ISM (HI and CO). Using these power law exponents, one can define a $Q$ stability parameter from Eq.\,\ref{eq:disprel2} or \ref{eq:coupleddisp} that quantifies the threshold for instability, in contrast to the case where the disc fragments due to the lack of stellar feedback support.

\section{Discussion and Conclusions}
\label{sect:discussion}
Gravitational instabilities are thought to play an important role in galaxy evolution and star formation. In this work we have investigated, using numerical simulations of Milky Way-like disc galaxies, the nature of turbulence in the ISM and how this affects the gravitational stability of galaxies. Clumpy/turbulent discs are dynamically similar to gas discs with scale-dependent surface densities and velocity dispersions, i.e. $\Sigma(\ell)\sim \ell^a$ and $\sigma(\ell)\sim l^b$ respectively, where $\ell$ is the physical scale. By taking these ``turbulent'' scaling relations into account in the disc stability analysis, a wide variety of non-classical stability properties arise \citep{Romeo2010}. 

In order to quantify the kind of turbulent scaling that develop in the ISM of disc galaxies we used numerical simulations of multi-component, Milky Way-like, galactic discs. We studied two markedly different evolutionary scenarios using the same initial condition: (1) no stellar feedback, leading to complete disc fragmentation and (2) efficient stellar feedback leading to driven ISM turbulence. By accounting for the measured turbulent scalings in the stability analysis, we could more robustly characterize the level stability as a function of physical scale. Our key results are summarized below:

\begin{enumerate}
\item Our two different models of galaxy evolution lead to discs with nearly identical stability properties when quantified on $\kpc$-scales by the classical Toomre $Q$ for stars and gas separately ($Q\gg1$), as well as jointly ($Q_{\rm stars + gas}>1$, accounting for disc thickness); the discs are stable at all radii. This notion is in good agreement with observations of nearby disc galaxies, e.g. THINGS \citep{Leroy08} where all discs, smoothed on $\kpc$ scales, featured values of $Q_{\rm gas}$ and $Q_{\star}$ well above unity \citep[see also][]{RomeoWiegert2011}, raising doubts on the role of gravitational instabilities in star formation.

\item The two models feature markedly different average $\Sigma(\ell)$ on scales $\ell < 1\kpc$, with a steepening of $\Sigma(\ell)$ into $\sim\ell^{-1}$ in the fragmented disc and $\sim \ell^{0-1/3}$ for the feedback regulated disc. Less of a difference was found for $\sigma(\ell)$, with a Larson-like scaling \citep[$\sigma(\ell)\sim\ell^{1/2}$, ][]{larson81,Solomon1987} in both models. 

\item Introducing scale dependent variables in the multi-component stability analysis leads to a more robust characterization of the level of instability.  The feedback driven model is, on average, stable or marginally stable on all scales, in contrast to the model without feedback, for which we can clearly identify the  scales ($\ell\lesssim100\pc$) where gravitational instability, leading to cloud formation, typically occurs. Large scales ($\ell\gtrsim1\kpc$) show almost identical stability properties in both models, as the surface density and velocity dispersion ``saturate'' into more well-defined, non-turbulent quantities, explaining the similarity between the markedly different ISM models when adopting a traditional large scale stability analysis.

\item The disc stabilized by stellar feedback can still be, in a statistical sense,  well described by a Toomre-like $Q$ stability threshold. This is no longer true for the violently fragmenting disc, which enters a regime where a $Q$-like parameter loses its meaning and small scales are asymptotically more and more unstable, and stability can only occur on large scales \citep[see also][]{Romeo2010}.
\end{enumerate}

\cite{Hopkins2013turb} investigated gravitational fragmentation in turbulent media, arguing that turbulent flows are always unstable on some scale, given enough time, as a broad spectrum of stochastic density fluctuations exists that can produce rare, but gravitationally unstable regions. This notion is compatible with our results; by defining the typical densities and velocities existing at some scale in a mass weighted sense, we are biasing ourselves towards dense regions. This gives us the \emph{typical} structures, by mass, that exists in the ISM. However, the scatter in the dispersion relation for the model with feedback-driven turbulence (Fig.\,\ref{fig:w2}) can be substantial on small scales at any time, due to the wide range of densities in supersonically turbulent flows \citep[e.g.][]{Padoan1997,maclow:review04,kritsuk07}. This means parts of the disc will be unstable, although the corresponding mass in that component may not necessarily be dominant.

In this work we have only considered Milky Way-like galaxies, with gas fractions typical of $z=0$ discs. In high redshift counterparts, with gas fractions several times greater \citep{Tacconi2010}, turbulence may play an even greater role in shaping the ISM. Indeed, the observed high levels of turbulence \citep[e.g.][]{Forster2009} is thought to be responsible for the highly clumpy morphologies observed \citep[e.g.][]{Elmegreen09,Tacconi2010,Agertz09b,DSC09,Genzel2011,Romeo2014}. Current astronomical facilities such as ALMA can resolve the scaling properties of galactic turbulence in the cold molecular gas of high redshift systems, hence revealing the interplay between gravitational instability and turbulence in more extreme environments. 

It is important to consider the limitations and assumptions behind the analysis presented in this work. The dispersion relation (Eq.\,\ref{eq:disprel1} and \ref{eq:disprel2}) is formally only describing stability against local axisymmetric perturbations, i.e. it assumes that $kR\gg1$. This is the short-wavelength approximation and is satisfied in our work as we only carry out our analysis at $R=4 \kpc$, and with $k=2\pi/\ell$ this holds on all scales we have considered. However, the assumption of axisymmetry is not true in general, and nonaxisymmetric perturbations are thought to have a greater destabilizing effect, i.e. discs that are formally stable ($Q>1$) can be locally unstable against such perturbations. Local nonaxisymmetric stability criteria are far more complex than Toomre's criterion as they depend critically on how tightly wound the perturbations are, and any such criteria cannot, in general, be expressed in terms of a single parameter akin to Toomre's $Q$ \citep[e.g.][]{LauBertin1978,Bertin1989,Jog1992,GrivGedalin2012}. We leave an analysis accounting for more general perturbations for future work.

In future work (Grisdale et al. in prep) we will extend the analysis presented in this paper to quantify the gravitational stability for different gas phases, regions of the disc, as a function of local turbulence driving strength etc., and how this connects to properties of observed galaxies.

\section*{acknowledgments}
We thank the referee for constructive comments and suggestions.

\footnotesize{
\bibliographystyle{mn2e}
\bibliography{ms.bbl}

\begin{thebibliography}{104}
\providecommand{\natexlab}[1]{#1}

\bibitem[{{Agertz} \& {Kravtsov}(2014)}]{AgertzKravtsov2014}
{Agertz} O., {Kravtsov} A.~V., 2014, ArXiv e-prints, 1404.2613

\bibitem[{{Agertz} et~al.(2009{\natexlab{a}}){Agertz}, {Lake}, {Teyssier},
  {Moore}, {Mayer} \& {Romeo}}]{Agertz09}
{Agertz} O., {Lake} G., {Teyssier} R., {Moore} B., {Mayer} L., {Romeo} A.~B.,
  2009{\natexlab{a}}, \mnras, 392, 294

\bibitem[{{Agertz} et~al.(2009{\natexlab{b}}){Agertz}, {Teyssier} \&
  {Moore}}]{Agertz09b}
{Agertz} O., {Teyssier} R., {Moore} B., 2009{\natexlab{b}}, \mnras, 397, L64

\bibitem[{{Agertz} et~al.(2013){Agertz}, {Kravtsov}, {Leitner} \&
  {Gnedin}}]{Agertz2013}
{Agertz} O., {Kravtsov} A.~V., {Leitner} S.~N., {Gnedin} N.~Y., 2013, \apj,
  770, 25

\bibitem[{{Ballesteros-Paredes} et~al.(2011){Ballesteros-Paredes}, {Hartmann},
  {V{\'a}zquez-Semadeni}, {Heitsch} \&
  {Zamora-Avil{\'e}s}}]{Ballesteros-Paredes2011}
{Ballesteros-Paredes} J., {Hartmann} L.~W., {V{\'a}zquez-Semadeni} E.,
  {Heitsch} F., {Zamora-Avil{\'e}s} M.~A., 2011, \mnras, 411, 65

\bibitem[{{Beaumont} et~al.(2012){Beaumont}, {Goodman}, {Alves}, {Lombardi},
  {Rom{\'a}n-Z{\'u}{\~n}iga}, {Kauffmann} \& {Lada}}]{Beaumont2012}
{Beaumont} C.~N., {Goodman} A.~A., {Alves} J.~F., {Lombardi} M.,
  {Rom{\'a}n-Z{\'u}{\~n}iga} C.~G., {Kauffmann} J., {Lada} C.~J., 2012, \mnras,
  423, 2579

\bibitem[{{Begelman} \& {Shlosman}(2009)}]{Begelman2009}
{Begelman} M.~C., {Shlosman} I., 2009, \apjl, 702, L5

\bibitem[{{Begum} et~al.(2006){Begum}, {Chengalur} \& {Bhardwaj}}]{Begum2006}
{Begum} A., {Chengalur} J.~N., {Bhardwaj} S., 2006, \mnras, 372, L33

\bibitem[{{Bertin} \& {Romeo}(1988)}]{bertinromeo88}
{Bertin} G., {Romeo} A.~B., 1988, \aap, 195, 105

\bibitem[{{Bertin} et~al.(1989){Bertin}, {Lin}, {Lowe} \&
  {Thurstans}}]{Bertin1989}
{Bertin} G., {Lin} C.~C., {Lowe} S.~A., {Thurstans} R.~P., 1989, \apj, 338, 104

\bibitem[{{Binney} \& {Tremaine}(2008)}]{BinneyTremaine2008}
{Binney} J., {Tremaine} S., 2008, {Galactic Dynamics: Second Edition}.
  Princeton University Press

\bibitem[{{Block} et~al.(2010){Block}, {Puerari}, {Elmegreen} \&
  {Bournaud}}]{Block2010}
{Block} D.~L., {Puerari} I., {Elmegreen} B.~G., {Bournaud} F., 2010, \apjl,
  718, L1

\bibitem[{{Blondin} et~al.(1998){Blondin}, {Wright}, {Borkowski} \&
  {Reynolds}}]{Blondin1998}
{Blondin} J.~M., {Wright} E.~B., {Borkowski} K.~J., {Reynolds} S.~P., 1998,
  \apj, 500, 342

\bibitem[{{Bolatto} et~al.(2008){Bolatto}, {Leroy}, {Rosolowsky}, {Walter} \&
  {Blitz}}]{bolatto_etal08}
{Bolatto} A.~D., {Leroy} A.~K., {Rosolowsky} E., {Walter} F., {Blitz} L., 2008,
  \apj, 686, 948

\bibitem[{{Bournaud} \& {Elmegreen}(2009)}]{Bournaud09}
{Bournaud} F., {Elmegreen} B.~G., 2009, \apjl, 694, L158

\bibitem[{{Bournaud} et~al.(2010){Bournaud}, {Elmegreen}, {Teyssier}, {Block}
  \& {Puerari}}]{Bournaud2010}
{Bournaud} F., {Elmegreen} B.~G., {Teyssier} R., {Block} D.~L., {Puerari} I.,
  2010, \mnras, 409, 1088

\bibitem[{{Cald{\'u}-Primo} et~al.(2013)}]{CalduPrimo2013}
{Cald{\'u}-Primo} A., {Schruba} A., {Walter} F., {Leroy} A., {Sandstrom} K.,
  {de Blok} W.~J.~G., {Ianjamasimanana} R., {Mogotsi} K.~M., 2013, \aj, 146,
  150

\bibitem[{{Cioffi} et~al.(1988){Cioffi}, {McKee} \&
  {Bertschinger}}]{Cioffi1988}
{Cioffi} D.~F., {McKee} C.~F., {Bertschinger} E., 1988, \apj, 334, 252

\bibitem[{{Collins} et~al.(2012){Collins}, {Kritsuk}, {Padoan}, {Li}, {Xu},
  {Ustyugov} \& {Norman}}]{Collins2012}
{Collins} D.~C., {Kritsuk} A.~G., {Padoan} P., {Li} H., {Xu} H., {Ustyugov}
  S.~D., {Norman} M.~L., 2012, \apj, 750, 13

\bibitem[{{Combes} et~al.(2012){Combes}, {Boquien}, {Kramer} \&
  et~al.}]{Combes2012}
{Combes} F., {Boquien} M., {Kramer} C., et~al., 2012, \aap, 539, A67

\bibitem[{{Dekel} et~al.(2009){Dekel}, {Sari} \& {Ceverino}}]{DSC09}
{Dekel} A., {Sari} R., {Ceverino} D., 2009, ArXiv e-prints

\bibitem[{{Dobbs}(2014)}]{Dobbs2014}
{Dobbs} C.~L., 2014, ArXiv e-prints

\bibitem[{{Dutta} et~al.(2008){Dutta}, {Begum}, {Bharadwaj} \&
  {Chengalur}}]{Dutta2008}
{Dutta} P., {Begum} A., {Bharadwaj} S., {Chengalur} J.~N., 2008, \mnras, 384,
  L34

\bibitem[{{Dutta} et~al.(2009){Dutta}, {Begum}, {Bharadwaj} \&
  {Chengalur}}]{Dutta2009}
{Dutta} P., {Begum} A., {Bharadwaj} S., {Chengalur} J.~N., 2009, \mnras, 397,
  L60

\bibitem[{{Dutta} et~al.(2013){Dutta}, {Begum}, {Bharadwaj} \&
  {Chengalur}}]{Dutta2013}
{Dutta} P., {Begum} A., {Bharadwaj} S., {Chengalur} J.~N., 2013, New Astronomy,
  19, 89

\bibitem[{{Elmegreen}(1995)}]{Elmegreen1995}
{Elmegreen} B.~G., 1995, \mnras, 275, 944

\bibitem[{{Elmegreen}(1996)}]{Elmegreen1996}
{Elmegreen} B.~G., 1996, in D.L. {Block}, J.M. {Greenberg}, eds, New
  Extragalactic Perspectives in the New South Africa. Astrophysics and Space
  Science Library, Vol. 209, p. 467

\bibitem[{{Elmegreen} \& {Scalo}(2004)}]{elmegreen04}
{Elmegreen} B.~G., {Scalo} J., 2004, \araa, 42, 211

\bibitem[{{Elmegreen} et~al.(2001){Elmegreen}, {Kim} \&
  {Staveley-Smith}}]{ElmegreenLMC2001}
{Elmegreen} B.~G., {Kim} S., {Staveley-Smith} L., 2001, \apj, 548, 749

\bibitem[{{Elmegreen} et~al.(2009){Elmegreen}, {Elmegreen}, {Fernandez} \&
  {Lemonias}}]{Elmegreen09}
{Elmegreen} B.~G., {Elmegreen} D.~M., {Fernandez} M.~X., {Lemonias} J.~J.,
  2009, \apj, 692, 12

\bibitem[{{Federrath}(2013)}]{Federrath2013}
{Federrath} C., 2013, \mnras, 436, 1245

\bibitem[{{Federrath} \& {Klessen}(2012)}]{Federrath2012}
{Federrath} C., {Klessen} R.~S., 2012, \apj, 761, 156

\bibitem[{{Federrath} et~al.(2010){Federrath}, {Roman-Duval}, {Klessen},
  {Schmidt} \& {Mac Low}}]{Federrath2010}
{Federrath} C., {Roman-Duval} J., {Klessen} R.~S., {Schmidt} W., {Mac Low}
  M.~M., 2010, \aap, 512, A81

\bibitem[{{Fleck}(1996)}]{Fleck1996}
{Fleck} Jr. R.~C., 1996, \apj, 458, 739

\bibitem[{{F{\"o}rster Schreiber} et~al.(2009)}]{Forster2009}
{F{\"o}rster Schreiber} N.~M. et~al., 2009, \apj, 706, 1364

\bibitem[{{Fujimoto} et~al.(2014){Fujimoto}, {Tasker}, {Wakayama} \&
  {Habe}}]{Fujimoto2014}
{Fujimoto} Y., {Tasker} E.~J., {Wakayama} M., {Habe} A., 2014, \mnras, 439, 936

\bibitem[{{Gammie} et~al.(1991){Gammie}, {Ostriker} \& {Jog}}]{gammie91}
{Gammie} C.~F., {Ostriker} J.~P., {Jog} C.~J., 1991, \apj, 378, 565

\bibitem[{{Gatto} et~al.(2014)}]{Gatto2014}
{Gatto} A. et~al., 2014, ArXiv e-prints

\bibitem[{{Genzel} et~al.(2011){Genzel}, {Newman}, {Jones}, {F{\"o}rster
  Schreiber} \& {et al.}}]{Genzel2011}
{Genzel} R., {Newman} S., {Jones} T., {F{\"o}rster Schreiber} N.~M., {et al.},
  2011, \apj, 733, 101

\bibitem[{{Griv} \& {Gedalin}(2012)}]{GrivGedalin2012}
{Griv} E., {Gedalin} M., 2012, \mnras, 422, 600

\bibitem[{{Guillet} \& {Teyssier}(2011)}]{GuilletTeyssier2011}
{Guillet} T., {Teyssier} R., 2011, Journal of Computational Physics, 230, 4756

\bibitem[{{Hernquist}(1990)}]{Hernquist1990}
{Hernquist} L., 1990, \apj, 356, 359

\bibitem[{{Hernquist}(1993)}]{Hernquist1993}
{Hernquist} L., 1993, \apjs, 86, 389

\bibitem[{{Heyer} et~al.(2009){Heyer}, {Krawczyk}, {Duval} \&
  {Jackson}}]{Heyer2009}
{Heyer} M., {Krawczyk} C., {Duval} J., {Jackson} J.~M., 2009, \apj, 699, 1092

\bibitem[{{Hockney} \& {Eastwood}(1981)}]{Hockney1981}
{Hockney} R.~W., {Eastwood} J.~W., 1981, {Computer Simulation Using Particles}.
  New York: McGraw-Hill, 1981

\bibitem[{{Hoffmann} \& {Romeo}(2012)}]{HoffmannRomeo2012}
{Hoffmann} V., {Romeo} A.~B., 2012, \mnras, 425, 1511

\bibitem[{{Hopkins}(2012)}]{HopkinsGMC2012}
{Hopkins} P.~F., 2012, \mnras, 423, 2016

\bibitem[{{Hopkins} \& {Christiansen}(2013)}]{Hopkins2013turb}
{Hopkins} P.~F., {Christiansen} J.~L., 2013, \apj, 776, 48

\bibitem[{{Hopkins} et~al.(2012){Hopkins}, {Quataert} \&
  {Murray}}]{Hopkins2012structure}
{Hopkins} P.~F., {Quataert} E., {Murray} N., 2012, \mnras, 421, 3488

\bibitem[{{Hughes} et~al.(2010)}]{Hughes2010}
{Hughes} A. et~al., 2010, \mnras, 406, 2065

\bibitem[{{Jog}(1992)}]{Jog1992}
{Jog} C.~J., 1992, \apj, 390, 378

\bibitem[{{Jog}(1996)}]{Jog1996}
{Jog} C.~J., 1996, \mnras, 278, 209

\bibitem[{{Jog} \& {Ostriker}(1988)}]{jogostriker88}
{Jog} C.~J., {Ostriker} J.~P., 1988, \apj, 328, 404

\bibitem[{{Jog} \& {Solomon}(1984)}]{JogSolomon1984}
{Jog} C.~J., {Solomon} P.~M., 1984, \apj, 276, 114

\bibitem[{{Kauffmann} et~al.(2010){Kauffmann}, {Pillai}, {Shetty}, {Myers} \&
  {Goodman}}]{Kauffmann2010}
{Kauffmann} J., {Pillai} T., {Shetty} R., {Myers} P.~C., {Goodman} A.~A., 2010,
  \apj, 716, 433

\bibitem[{{Kim} \& {Ostriker}(2014)}]{KimOstriker2014}
{Kim} C.~G., {Ostriker} E.~C., 2014, ArXiv e-prints

\bibitem[{{Kim} et~al.(2014){Kim}, {Abel} \& {Agertz}}]{agora}
{Kim} J.~h., {Abel} T., {Agertz} O.~e.~a., 2014, \apjs, 210, 14

\bibitem[{{Kim} et~al.(2007)}]{Kim2007}
{Kim} S. et~al., 2007, \apjs, 171, 419

\bibitem[{{Kowal} \& {Lazarian}(2007)}]{Kowal2007}
{Kowal} G., {Lazarian} A., 2007, \apjl, 666, L69

\bibitem[{{Kritsuk} et~al.(2007){Kritsuk}, {Norman}, {Padoan} \&
  {Wagner}}]{kritsuk07}
{Kritsuk} A.~G., {Norman} M.~L., {Padoan} P., {Wagner} R., 2007, \apj, 665, 416

\bibitem[{{Kritsuk} et~al.(2013){Kritsuk}, {Lee} \& {Norman}}]{Kritsuk2013}
{Kritsuk} A.~G., {Lee} C.~T., {Norman} M.~L., 2013, \mnras, 436, 3247

\bibitem[{{Larson}(1981)}]{larson81}
{Larson} R.~B., 1981, \mnras, 194, 809

\bibitem[{{Lau} \& {Bertin}(1978)}]{LauBertin1978}
{Lau} Y.~Y., {Bertin} G., 1978, \apj, 226, 508

\bibitem[{{Lazarian} \& {Pogosyan}(2000)}]{LazarianPogosyan2000}
{Lazarian} A., {Pogosyan} D., 2000, \apj, 537, 720

\bibitem[{{Leroy} et~al.(2008){Leroy}, {Walter}, {Brinks}, {Bigiel}, {de Blok},
  {Madore} \& {Thornley}}]{Leroy08}
{Leroy} A.~K., {Walter} F., {Brinks} E., {Bigiel} F., {de Blok} W.~J.~G.,
  {Madore} B., {Thornley} M.~D., 2008, \aj, 136, 2782

\bibitem[{{Lin} \& {Shu}(1966)}]{LinShu1966}
{Lin} C.~C., {Shu} F.~H., 1966, Proceedings of the National Academy of Science,
  55, 229

\bibitem[{{Lombardi} et~al.(2010){Lombardi}, {Lada} \& {Alves}}]{Lombardi2010}
{Lombardi} M., {Lada} C.~J., {Alves} J., 2010, \aap, 512, A67

\bibitem[{{Mac Low} \& {Klessen}(2004)}]{maclow:review04}
{Mac Low} M.~M., {Klessen} R.~S., 2004, Reviews of Modern Physics, 76, 125

\bibitem[{{Martizzi} et~al.(2014){Martizzi}, {Faucher-Gigu{\`e}re} \&
  {Quataert}}]{Martizzi2014}
{Martizzi} D., {Faucher-Gigu{\`e}re} C.~A., {Quataert} E., 2014, ArXiv e-prints

\bibitem[{{McKee} \& {Ostriker}(2007)}]{mckeeostriker07}
{McKee} C.~F., {Ostriker} E.~C., 2007, \araa, 45, 565

\bibitem[{{Navarro} et~al.(1996){Navarro}, {Frenk} \& {White}}]{nfw1996}
{Navarro} J.~F., {Frenk} C.~S., {White} S.~D.~M., 1996, \apj, 462, 563

\bibitem[{{Padoan} \& {Nordlund}(2011)}]{PadoanNordlund2011}
{Padoan} P., {Nordlund} {\AA}., 2011, \apj, 730, 40

\bibitem[{{Padoan} et~al.(1997){Padoan}, {Jones} \& {Nordlund}}]{Padoan1997}
{Padoan} P., {Jones} B.~J.~T., {Nordlund} A.~P., 1997, \apj, 474, 730

\bibitem[{{Price} \& {Federrath}(2010)}]{PriceFederrath2010}
{Price} D.~J., {Federrath} C., 2010, \mnras, 406, 1659

\bibitem[{{Rafikov}(2001)}]{rafikov01}
{Rafikov} R.~R., 2001, \mnras, 323, 445

\bibitem[{{Renaud} et~al.(2013)}]{Renaud2013}
{Renaud} F. et~al., 2013, \mnras, 436, 1836

\bibitem[{{Roman-Duval} et~al.(2010){Roman-Duval}, {Jackson}, {Heyer},
  {Rathborne} \& {Simon}}]{Roman-Duval2010}
{Roman-Duval} J., {Jackson} J.~M., {Heyer} M., {Rathborne} J., {Simon} R.,
  2010, \apj, 723, 492

\bibitem[{{Romeo}(1992)}]{romeo92}
{Romeo} A.~B., 1992, \mnras, 256, 307

\bibitem[{{Romeo}(1994)}]{romeo94}
{Romeo} A.~B., 1994, \aap, 286, 799

\bibitem[{{Romeo} \& {Agertz}(2014)}]{Romeo2014}
{Romeo} A.~B., {Agertz} O., 2014, \mnras, 442, 1230

\bibitem[{{Romeo} \& {Falstad}(2013)}]{RomeoFalstad2013}
{Romeo} A.~B., {Falstad} N., 2013, \mnras, 433, 1389

\bibitem[{{Romeo} \& {Wiegert}(2011)}]{RomeoWiegert2011}
{Romeo} A.~B., {Wiegert} J., 2011, \mnras, 416, 1191

\bibitem[{{Romeo} et~al.(2008){Romeo}, {Agertz}, {Moore} \& {Stadel}}]{Romeo08}
{Romeo} A.~B., {Agertz} O., {Moore} B., {Stadel} J., 2008, \apj, 686, 1

\bibitem[{{Romeo} et~al.(2010){Romeo}, {Burkert} \& {Agertz}}]{Romeo2010}
{Romeo} A.~B., {Burkert} A., {Agertz} O., 2010, \mnras, 407, 1223

\bibitem[{{Roy} et~al.(2008){Roy}, {Peedikakkandy} \& {Chengalur}}]{Roy2008}
{Roy} N., {Peedikakkandy} L., {Chengalur} J.~N., 2008, \mnras, 387, L18

\bibitem[{{S{\'a}nchez} et~al.(2010){S{\'a}nchez}, {A{\~n}ez}, {Alfaro} \&
  {Crone Odekon}}]{Sanchez2010}
{S{\'a}nchez} N., {A{\~n}ez} N., {Alfaro} E.~J., {Crone Odekon} M., 2010, \apj,
  720, 541

\bibitem[{{Schmidt} et~al.(2008){Schmidt}, {Federrath} \&
  {Klessen}}]{Schmidt2008}
{Schmidt} W., {Federrath} C., {Klessen} R., 2008, Physical Review Letters, 101,
  194505

\bibitem[{{Shadmehri} \& {Khajenabi}(2012)}]{Shadmehri2012}
{Shadmehri} M., {Khajenabi} F., 2012, \mnras, 421, 841

\bibitem[{{Shapiro} et~al.(2008)}]{Shapiro08}
{Shapiro} K.~L. et~al., 2008, \apj, 682, 231

\bibitem[{{Simpson} et~al.(2014){Simpson}, {Bryan}, {Hummels} \&
  {Ostriker}}]{Simpson2014}
{Simpson} C.~M., {Bryan} G.~L., {Hummels} C., {Ostriker} J.~P., 2014, ArXiv
  e-prints

\bibitem[{{Solomon} et~al.(1987){Solomon}, {Rivolo}, {Barrett} \&
  {Yahil}}]{Solomon1987}
{Solomon} P.~M., {Rivolo} A.~R., {Barrett} J., {Yahil} A., 1987, \apj, 319, 730

\bibitem[{{Springel}(2000)}]{Springel2000}
{Springel} V., 2000, \mnras, 312, 859

\bibitem[{{Springel} et~al.(2005){Springel}, {Di Matteo} \&
  {Hernquist}}]{SpringelMatteoHernquist2005}
{Springel} V., {Di Matteo} T., {Hernquist} L., 2005, \mnras, 361, 776

\bibitem[{{Stanimirovic} et~al.(1999){Stanimirovic}, {Staveley-Smith},
  {Dickey}, {Sault} \& {Snowden}}]{Stanimirovic1999}
{Stanimirovic} S., {Staveley-Smith} L., {Dickey} J.~M., {Sault} R.~J.,
  {Snowden} S.~L., 1999, \mnras, 302, 417

\bibitem[{{Swinbank} et~al.(2011)}]{Swinbank2011}
{Swinbank} A.~M. et~al., 2011, \apj, 742, 11

\bibitem[{{Tacconi} et~al.(2010)}]{Tacconi2010}
{Tacconi} L.~J. et~al., 2010, Nature, 463, 781

\bibitem[{{Tamburro} et~al.(2009)}]{Tamburro2009}
{Tamburro} D., {Rix} H.~W., {Leroy} A.~K., {Mac Low} M.~M., {Walter} F.,
  {Kennicutt} R.~C., {Brinks} E., {de Blok} W.~J.~G., 2009, \aj, 137, 4424

\bibitem[{{Tasker} \& {Tan}(2009)}]{tasker09}
{Tasker} E.~J., {Tan} J.~C., 2009, \apj, 700, 358

\bibitem[{{Teyssier}(2002)}]{teyssier02}
{Teyssier} R., 2002, \aap, 385, 337

\bibitem[{{Thornton} et~al.(1998){Thornton}, {Gaudlitz}, {Janka} \&
  {Steinmetz}}]{Thornton1998}
{Thornton} K., {Gaudlitz} M., {Janka} H.~T., {Steinmetz} M., 1998, \apj, 500,
  95

\bibitem[{{Toomre}(1964)}]{toomre64}
{Toomre} A., 1964, \apj, 139, 1217

\bibitem[{{Vandervoort}(1970)}]{Vandervoort1970}
{Vandervoort} P.~O., 1970, \apj, 161, 87

\bibitem[{{Walter} et~al.(2008){Walter}, {Brinks}, {de Blok}, {Bigiel},
  {Kennicutt}, {Thornley} \& {Leroy}}]{Walter2008}
{Walter} F., {Brinks} E., {de Blok} W.~J.~G., {Bigiel} F., {Kennicutt} Jr.
  R.~C., {Thornley} M.~D., {Leroy} A., 2008, \aj, 136, 2563

\bibitem[{{Zhang} et~al.(2012){Zhang}, {Hunter} \& {Elmegreen}}]{Zhang2012}
{Zhang} H.~X., {Hunter} D.~A., {Elmegreen} B.~G., 2012, \apj, 754, 29

\end{thebibliography}
}

\end{document}